\documentclass[letterpaper,12pt]{article}
\usepackage{color,amssymb,enumerate,float}
\usepackage{amsmath}

\usepackage{ifpdf}
\ifpdf
\usepackage[pdftex]{graphicx}
\usepackage[pdftex,unicode,implicit]{hyperref}

\hypersetup{
	pdftitle     = {}, 
	pdfkeywords  = {},
	pdfauthor    = {},
	pdfcreator   = {pdf\LaTeXe\ with package \flqq hyperref\frqq},
	pdfproducer  = {pdf\LaTeXe\ with package \flqq hyperref\frqq},
	pdfpagemode  = UseNone,  
	pdffitwindow = true,  
	unicode      = true,
	plainpages   = true,
	colorlinks   = true,  
	citecolor    = black,  
	urlcolor     = blue, 
	linkcolor    = black
}

\else

\usepackage[dvips]{graphicx}

\fi 

\makeatletter
\@addtoreset{equation}{section}
\makeatother

\begin{document}
	
\thispagestyle{empty}
	
\begin{center}
{\bf \LARGE 
Quasi-Einstein metrics from Ho\v{r}ava gravity
}
\vspace*{15mm}
		
{\large Jorge Bellorin}
\vspace{3ex}
		
{\it Department of Physics, Universidad de Antofagasta, 1240000 Antofagasta, Chile.}
				
\vspace*{2ex}
{\tt jorge.bellorin@uantof.cl} \hspace{1em}
		
\vspace*{15mm}
{\bf Abstract}
\begin{quotation}{\small
In this note we present a direct relation between the geometrical definition of a quasi-Einstein metric and Ho\v{r}ava gravity: the solutions of the large-distance effective action of Ho\v{r}ava gravity that are static and orthogonal are quasi-Einstein metrics. The converse is also true if the function over the quasi-Einstein manifold satisfies a given equation.
}
\end{quotation}
\end{center}

\section{Introduction}
Frequently, concepts of differential geometry and gravitation coincide. In this sense, a feature that is broadly studied is the condition of an Einstein metric, which, stated for a Riemannian metric, establishes the proportionality between the Ricci tensor and the metric,
\begin{equation}
 R_{ij} = k g_{ij} 
 \label{EinsteinCondition}
\end{equation}
(the notation with indices, assuming a given local coordinate basis, is convenient to connect with a field theory). In general relativity, dealing with a four-dimensional spacetime metric, this condition acquires a foundational interpretation, since it is the Einstein equations in vacuum with a cosmological constant $\Lambda$,
\begin{equation}
 R_{\mu\nu} - \Lambda g_{\mu\nu} = 0 \,.
\end{equation}

The definition of an Einstein metric (\ref{EinsteinCondition}) has been extended to the concept of  quasi-Einstein metrics \cite{Case:2011}, substituting the Ricci tensor by the $m$-Bakry-Emery Ricci tensor. Given the triplet $(M,g,f)$, where $M$ is a Riemannian manifold with metric $g$, and $f$ is a smooth real valued function on the manifold, the $m$-Bakry-Emery Ricci tensor ${R_f^m}_{ij}$ is defined by
\begin{equation}
	{R_f^m}_{ij} = 
 	R_{ij} + \nabla_{i j} f - \frac{1}{m} \nabla_i f \nabla_j f  
 	\quad
 	\text{for}
 	\quad
 	0 < m \leq \infty \,.
\end{equation}
This tensor becomes equal to the Ricci tensor for the case of constant $f$. A quasi-Einstein metric is defined by the condition of proportionality between $m$-Bakry-Emery Ricci tensor and the metric,
\begin{equation}
 	R_{ij} + \nabla_{i j} f - \frac{1}{m} \nabla_i f \nabla_j f 
 	= k g_{ij} \,,
 	\label{quasiEinsteindef}
\end{equation}
for some $k\in \mathbb{R}$. In the case $m\rightarrow \infty$, this equation is exactly the gradient Ricci soliton equation.

Quasi-Einstein metrics have been related with configurations in general relativity, see for example Refs.~\cite{Bahuaud:2022iao,Bahuaud:2023wsi,Colling:2024usk}. In this note, we present a relation that the quasi-Einstein metrics have with another gravitational theory: the Ho\v{r}ava gravity \cite{Horava:2009uw}. This theory is a proposal for a theory of quantum gravity that have strong evidence to be perturbatively renormalizable. The failure in the perturbative renormalization of the quantum version of general relativity is the problem that has motivated for a long time the study of alternative or complementary theories in the quantum domain. Another important feature of the Ho\v{r}ava gravity is its unitarity, which is a condition on the physical consistency of its quantum modes. The capacity of conciliating the two aspects, renormalizability and unitarity, gives interest to the Ho\v{r}ava theory as a proposal for quantum gravity.

The characteristic feature of the Ho\v{r}ava gravity is the breaking of the local Lorentz symmetry. In this way, the theory deals with a nonrelativistic Lagrangian. Although the theory was originally conceived as an ultraviolet completion of general relativity, certainly the nonrelativistic terms can be of the same scale as the terms of general relativity, that is, terms of the scale of the classical physics. Precisely these terms in the Lagrangian, which we call the large-distance effective Lagrangian of the Ho\v{r}ava gravity, are the terms that lead us to the connection with the quasi-Einstein metrics. We state the connection on a precise form: the static-orthogonal solutions of the large-distance effective action of the Ho\v{r}ava theory are quasi-Einstein metrics. The converse is also true if we impose an additional condition on the quasi-Einstein metrics, which can be interpreted as a condition on $f$ (an identification of the constants in both sides is also required in the converse case).


\section{Ho\v{r}ava theory and the static-orthogonal condition} 
The Ho\v{r}ava gravity is formulated assuming the breaking of the Local Lorentz symmetry. Indeed, the underlying geometric structure is not a pseudo-Riemannian manifold representing a spacetime, but a foliation of spacelike hypersurfaces along a given direction of time. It is appropriated to introduce the field variables characteristic of the Arnowitt-Deser-Misner (ADM) parametrization, which are the Riemannian metric $g_{ij}$ on each leaf of the foliation, the lapse function $N$ and the shift vector $N^i$. The gauge symmetry on these fields is given by the foliation-preserving diffeomorphisms group. The ADM fields in general depend parametrically on time, which means that they in general change when passing from a leaf to another. In the formulation of the Ho\v{r}ava gravity, there is a special case where $N$ can be regarded as a function only of time. We do not consider this case; hence we deal with what is called the nonprojectable Ho\v{r}ava gravity.

The Lagrangian of the Ho\v{r}ava gravity is of second order in time derivatives, but in spatial derivatives it grows from the second to higher orders. The terms of higher-order in spatial derivatives are the ones that improve the loop divergences at the ultraviolet. In this study we concentrate on the terms that are of second order in time and spatial derivatives. One can justify physically this truncation by arguing that the higher orders in derivatives are less important for the physics of large distances, where the quantum consistency is not under scrutiny. In this sense, we deal with the large-distance effective action.

The large-distance effective action of the complete, nonprojectable Ho\v{r}ava gravity is given by \cite{Horava:2009uw,Blas:2009qj}
\begin{equation}
 S = \int dt d^3x \sqrt{g} N 
 ( G^{ijkl} K_{ij} K_{kl} + \beta ( R - 2\Lambda ) + \alpha N^{-2} \nabla_i N \nabla^i N ),
 \label{lagrangianaction}
\end{equation}
where
\begin{eqnarray}
  &&
	G^{ijkl} = 
    \frac{1}{2} ( g^{ik} g^{jl} + g^{il} g^{jk} ) 
    - \lambda g^{ij} g^{kl} \,,
  \\&&
	K_{ij} = \frac{1}{2N} ( \dot{g}_{ij} - 2 \nabla_{(i} N_{j)} ) \,,
\end{eqnarray}
$\alpha,\beta,\lambda$ are coupling constants, and $\Lambda$ can be associated with a cosmological constant. We use geometric notation for the spatial metric: we raise and lower spatial indices with $g_{ij}$, $\nabla_i$ is the (torsionless) affine connection of $g_{ij}$, and $R_{ijk}{}^{l},R_{ij},R$ are the standard curvature tensors of $g_{ij}$. The field equations, which are derived by taking variations of (\ref{lagrangianaction}) with respect to $g_{ij}$, $N$ and $N_i$, are given, respectively, by
\begin{eqnarray}
	\frac{1}{\sqrt{g}} \frac{\partial}{\partial t} 
	( \sqrt{g} G^{ijkl} K_{kl} )
	+ 2 N ( K^{ik} K_k{}^j - \lambda K K^{ij} )
	- \frac{1}{2} N g^{ij} G^{klmn} K_{kl} K_{mn} 
	& & \nonumber \\
	+ 2 \nabla_k ( G^{kmn(i} K_{mn} N^{j)} )
	- \nabla_k ( G^{mnij} K_{mn} N^k )
	+ \beta N (R^{ij} - \frac{1}{2} g^{ij} R + g^{i j} \Lambda  )
	& & \nonumber \\ 
	- \beta ( \nabla^{i j} N - g^{ij} \nabla^2 N )
	+ \alpha N^{-1} ( \nabla^i N \nabla^j N 
	- \frac{1}{2} g^{ij} \nabla_k N \nabla^k N )
	& = & 0 \,,
	\label{einstein}
	\\
	G^{ijkl} K_{ij} K_{kl} - \beta ( R - 2\Lambda ) 
	+ 2 \alpha N^{-2} ( N \nabla^2 N - \frac{1}{2} \nabla_i N \nabla^i N ) 
	& = & 0 \,,
	\label{hamiltonianconstrainlagrange}
	\\
	\nabla_i ( G^{ijkl} K_{kl} ) &=& 0 \,.
	\label{momentunconstrainlagrange}
\end{eqnarray}
These field equations are to be interpreted as vacuum field equations since no source terms have been included in the Lagrangian (\ref{lagrangianaction}). The field equations (\ref{einstein}) -- (\ref{momentunconstrainlagrange}) have been used to study static and spherically symmetric solutions of the Ho\v{r}ava theory, see for example \cite{Kiritsis:2009vz,Bellorin:2014qca,Bellorin:2015oja, Bellorin:2025gtx}. The central result in this context (in three spatial dimensions) is that the static spherically symmetric solutions that naturally arise are spaces with a throat; that is, geometries with a minimum value of the radius, such that two spatial halves are connected by this throat. The same solutions were found previously in the context of the Einstein-aether theory \cite{Eling:2006df}, which is a covariant theory that, under some assumptions, becomes equivalent to the large-distance effective action (\ref{lagrangianaction}) of the Ho\v{r}ava gravity.

We define the static condition as the condition that all field variables are independent of time. The condition of orthogonality of the foliation is defined by $N^i=0$. We remark that, in Ho\v{r}ava theory, the gauge-symmetry group is different to the group of symmetries of general relativity; hence the managing of the shift vector $N^i$ is different. For this reason we impose the two conditions independently.

We evaluate the field equations (\ref{einstein}) -- (\ref{momentunconstrainlagrange}) on static-orthogonal configurations. A consequence of these conditions is $K_{ij} =0$. With this, the Eq.~(\ref{momentunconstrainlagrange}) is automatically solved. The Eqs.~(\ref{einstein}) and (\ref{hamiltonianconstrainlagrange}) reduce, respectively, to
\begin{eqnarray}
 &&
 \beta N (R^{ij} - \frac{1}{2} g^{ij} R + g^{ij} \Lambda )
 - \beta ( \nabla^{i j} N - g^{ij} \nabla^2 N )
 + \alpha N^{-1} ( \nabla^i N \nabla^j N 
 - \frac{1}{2} g^{ij} \nabla_k N \nabla^k N )
 = 0 \,, 
 \nonumber \\
 \label{eisnteinstatic}
 \\ &&
 \beta ( R - 2\Lambda ) 
 - \alpha ( 2 N^{-1} \nabla^2 N - N^{-2} \nabla_k N \nabla^k N ) 
 = 0 \,.
 \label{hamiltoniancstatic}
\end{eqnarray}
We perform some manipulation to bring these equations to simpler forms. The trace of Eq.~(\ref{eisnteinstatic}) yields
\begin{equation}
 \beta ( R - 6 \Lambda ) - 4 \beta N^{-1} \nabla^2 N
 + \alpha N^{-2} \nabla_k N \nabla^k N = 0 \,.
\end{equation}
A combination of this equation with Eq.~(\ref{hamiltoniancstatic}) yields
\begin{equation}
 \nabla^2 N = - \gamma N \,,
 \label{nablaN}
\end{equation}
where 
\begin{equation}
 \gamma \equiv 2\beta\Lambda/(2\beta - \alpha) \,.
 \label{gamma}
\end{equation}
By substituting this equation back in Eq.~(\ref{hamiltoniancstatic}), we obtain
\begin{equation}
 \beta R + \alpha N^{-2} \nabla_k N \nabla^k N  
 = - ( 3\alpha - 2\beta ) \gamma \,.
 \label{hamiltonianfin}
\end{equation}
Finally, the field equation (\ref{eisnteinstatic}), after substituting (\ref{nablaN}) and (\ref{hamiltonianfin}) into it, takes the form
\begin{equation}
	\beta R^{ij} - \beta N^{-1} \nabla^{i j} N 
	+ \alpha N^{-2} \nabla^i N \nabla^j N  = - ( \alpha - \beta ) \gamma g^{ij} \,.
	\label{eomstaticlag}
\end{equation}
The system of equations formed by the Eqs.~(\ref{nablaN}) and (\ref{eomstaticlag}) is equivalent to the field equations (\ref{eisnteinstatic}) and (\ref{hamiltoniancstatic}). Hence, the static-orthogonal solutions of the large-distance effective action of the Ho\v{r}ava gravity are determined by Eqs.~(\ref{nablaN}) and (\ref{eomstaticlag}).

We may change the lapse function by $N = e^{-f}$, such that the Eq.~(\ref{eomstaticlag}) becomes
\begin{equation}
	R_{ij} + \nabla_{i j} f - \frac{1}{m} \nabla_i f \nabla_j f  = 
	\frac{\gamma}{m} g_{ij} \,,
	\label{quasiEinstein}
\end{equation}
where 
\begin{equation}
m = ( 1 - \alpha/\beta )^{-1} \,.
\label{m}
\end{equation}
The equation (\ref{quasiEinstein}) is identical to the condition of a quasi-Einstein metric (\ref{quasiEinsteindef}); hence, we have proven that any static-orthogonal solution $(M, g, f)$ of the large-distance effective action of the Ho\v{r}ava gravity is a quasi-Einstein metric. The parameter $m$ of the $m$-Bakry-Emery Ricci tensor and the constant of proportionality $\gamma/m$ are given in terms of Ho\v{r}ava gravity's constants $\alpha$, $\beta$ and $\Lambda$. In the case $\Lambda = 0$, the theory requires that the $m$-Bakry-Emery tensor vanishes.

The equation (\ref{nablaN}) takes the form
\begin{equation}
  \nabla^2 f - \nabla_k f \nabla^k f = \gamma \,.
  \label{eqf}
\end{equation}
Therefore, the inverse relation works in the following way: let $(M, g, f)$ a quasi-Einstein metric, such that the constants are related to Ho\v{r}ava gravity's constants by the relations (\ref{gamma}) and (\ref{m}). If the function $f$ satisfies the Eq.~(\ref{eqf}), then the quasi-Einstein metric is a static-orthogonal solution of the large-distance effective action of the Ho\v{r}ava gravity\footnote{Of course, the two equations (\ref{quasiEinstein}) and (\ref{eqf}) define a system of equations for the pair $(g,f)$; not necessarily separated for $g$ and $f$, as we have presented the discussion in the main text for simplicity.}

As a final comment, we would like to highlight the conceptual similarities between the cases of general relativity and Ho\v{r}ava gravity. The condition of an Einstein metric is understood as the geometrical interpretation of the vacuum Einstein equations with cosmological constant, extended to the domain of Riemannian metrics. Here, we have found that the condition of quasi-Einstein metrics can be interpreted, under the static-orthogonal condition, as the vacuum field equation of another gravitational theory. The cosmological constant is also necessary to establish the proportionality between the tensor of interest and the metric.


\end{document}